\documentclass[english,twocolumn,amsmath,amssymb,nofootinbib,superscriptaddress,showpacs]{revtex4}

\usepackage[active]{srcltx}

\usepackage{amsfonts}
\usepackage{amssymb}
\usepackage{latexsym}
\usepackage{graphicx}
\usepackage{epsfig,subfigure}
\usepackage{epstopdf}
\usepackage[active]{srcltx}
\makeatother

\begin{document}

%\title{Data-adaptive unfolding of eigenspectra ensembles}

\title{Random Matrix  Spectra as a Time Series
%: \\ An approach with Singular Value Decomposition (SVD)
}

\author{R. Fossion\footnote{Email: fossion@nucleares.unam.mx}}
\affiliation{Instituto Nacional de Geriatr\'ia, Perif\'erico Sur No. 2767, 10200 M\'exico D.F., Mexico}
\affiliation{Centro de Ciencias de la Complejidad (C3),
Universidad Nacional Aut\'{o}noma de M\'{e}xico, 04510 M\'{e}xico D.F., Mexico}

\author{G. Torres Vargas}
\affiliation{Posgrado en Ciencias F\'isicas,
Universidad Nacional Aut\'{o}noma de M\'{e}xico, 04510 M\'{e}xico D.F., Mexico}

\author{J.C.~L\'{o}pez Vieyra}
\affiliation{Instituto de Ciencias Nucleares,
Universidad Nacional Aut\'{o}noma de M\'{e}xico, 04510 M\'{e}xico D.F., Mexico}

%\keywords{}

\begin{abstract}
\noindent
Spectra of ordered eigenvalues of finite Random Matrices are interpreted as a time series. Data-adaptive techniques from signal analysis are applied to decompose the spectrum in clearly differentiated trend and fluctuation modes, avoiding possible artifacts introduced by standard unfolding techniques. The fluctuation modes are scale invariant and follow different power laws for Poisson and Gaussian ensembles, which already during the unfolding allows to distinguish the two cases.   
\end{abstract}


\maketitle

\noindent The study of spectral fluctuations within the framework of Random Matrix Theory (RMT) is a standard tool in the statistical study of quantum chaos in the excitation spectra of quantum systems \cite{bro81,meh91,haa10,gom11}. Recently, the approach has found new applications in many fields, such as in the study of eigenspectra of adjacency matrices of networks \cite{luo06,jal09,bin10}, and eigenspectra of empirical correlation matrices in finance \cite{lal99,ple99,ple02}, the climate \cite{san01}, electro- and magnetoencephalography \cite{kwa00,seb03,mul06}, and in complex systems \cite{kwa12}. The interest of the approach lies in the fact that the level density fluctuations $\widetilde{\rho}(E)=\rho(E)-\overline{\rho}(E)$ around the smooth global density $\overline{\rho}(E)$ are universal and indicate the underlying symmetry class of the system \cite{meh91,erd12}. On the other hand, the global level density $\overline{\rho}(E)$ is system dependent, and an unfolding procedure needs to be performed, to separate the global and the fluctuating parts \cite{bro81}. The unfolding is straightforward if an analytical formula is known to describe the global level density $\overline{\rho}(E)$ for the system under study, such as e.g. the gaussian and semicircle distributions for Possion and GOE matrix ensembles from RMT \cite{meh91}, or the Marchenko-Pastur distribution for the Laguerre ensemble of random Wishart correlation matrices \cite{mar67}. However, such analytical formulae are formally only adequate in the asymptotic limit for spectra with an infinite number of levels. Often, an analytical form for $\overline{\rho}(E)$ is unknown, as is the case for adjacency matrices \cite{jal09}. In practical cases, having finite, {\it albeit} large matrices, the usual approach is then to project the sequence of ordered eigenvalues into unfolded values $E(n) \rightarrow \overline{\mathcal{N}} \left[ E(n) \right]$, using a smooth (often polynomial) approximation $\overline{\mathcal{N}}( E )$ to the accumulated density (step) function  $\mathcal{N}(E) = \int_{-\infty}^E \rho(E') dE'$ \cite{rel02,bro81,jal09}. After unfolding, the short-range and long-range correlations can be quantified using  standard fluctuations measures such as the Nearest-Neighbour Spacing (NNS) distribution, number variance $\Sigma^2$ and $\Delta_3$. In a recent approach, the unfolded fluctuations of the accumulated level density function $\widetilde{\mathcal{N}}(E) = \mathcal{N}(E) - \overline{\mathcal{N}}(E)$ (also called $\delta_n$ function) were interpreted as a time series  \cite{gom11,rel02, fal04}. This treatment opened the field to the application of specialized techniques from signal analysis, such as Fourier spectral analysis \cite{gom11,rel02,fal04, bin10}, Detrended Fluctuation Analysis (DFA) \cite{pen94,san06,lan08}, wavelets \cite{mal07}, Empirical Mode Decomposition (EMD) \cite{lan11,mor11,lan13}, and normal-mode analysis \cite{and99,jac01}. The result of these investigations is that for Gaussian RMT ensembles, the fluctuation time series is scale invariant (fractal), which in the Fourier power spectrum is reflected in a power law, 
\begin{equation}
P(f) \propto 1/f^\beta,
\label{EqFourier}
\end{equation}
where $f$ is the frequency of the periodic modes in which the time series is decomposed, whereas when more general non-periodic normal modes are used, a ``generalized power spectrum'' or so-called ``scree diagram'' results, 
 \begin{equation}
\lambda_k \propto 1/k^\gamma,
\label{EqScree}
\end{equation}
where $k$ is the index of the normal modes, and where $\beta=\gamma=2$ (Poisson limit) and $\beta=\gamma=1$ (GOE limit), such that the power law does not seem to depend on the basis used to decompose the time series \cite{gao03}. All fluctuation measures mentioned, are calculated after the prior technical step of the unfolding of the original eigenvalues. However, the statistical results can be quite sensitive to the specific unfolding procedure used (see e.g. \cite{gom02,abu12}). In signal analysis, a similar problem is how to define the \emph{trend} of non-stationary time series. It was concluded that the trend is an \emph{intrinsic} property of the time series that should not be defined by an external observer but should be obtained \emph{data-adaptively} from the data itself \cite{wu07}. The purpose of the present contribution is twofold: first, we propose to interpret the spectrum of original eigenvalues $E(n)$ directly as a time series, such that data-adaptive techniques from signal analysis can be used to decompose the sequence in a global and local part,
\begin{equation}
E(n) = \overline{E}(n)+\widetilde{E}(n),
\label{EqUnfolding}
\end{equation}
secondly, we will present one particular method with which this unfolding can be realized. We will see that the power law of eq.~(\ref{EqScree}) is obtained already during the proposed data-adaptive unfolding procedure.

\begin{figure}[htb!]
%%%%%%%%%%%%%%%%%%%%%%%%%%%%%%%%%%%%%
\begin{minipage}{.99\linewidth}    % START FRAME MINIPAGE
%%%%%%%%%%%%%%%%%%%%%%%%%%%%%%%%%%%%%
     %%%%%%%%%%%%%%%%%%%%%%%%%%%%%%%%%%%%%
     \begin{minipage}{.99\linewidth}    % START FRAME MINIPAGE
     %%%%%%%%%%%%%%%%%%%%%%%%%%%%%%%%%%%%%
        %%%%%%%%%%% FIG1A %%%%%%%%%%%%%%%%%%%
        \begin{minipage}{.49\linewidth}
        \begin{center}
        \includegraphics[width=.99\linewidth]{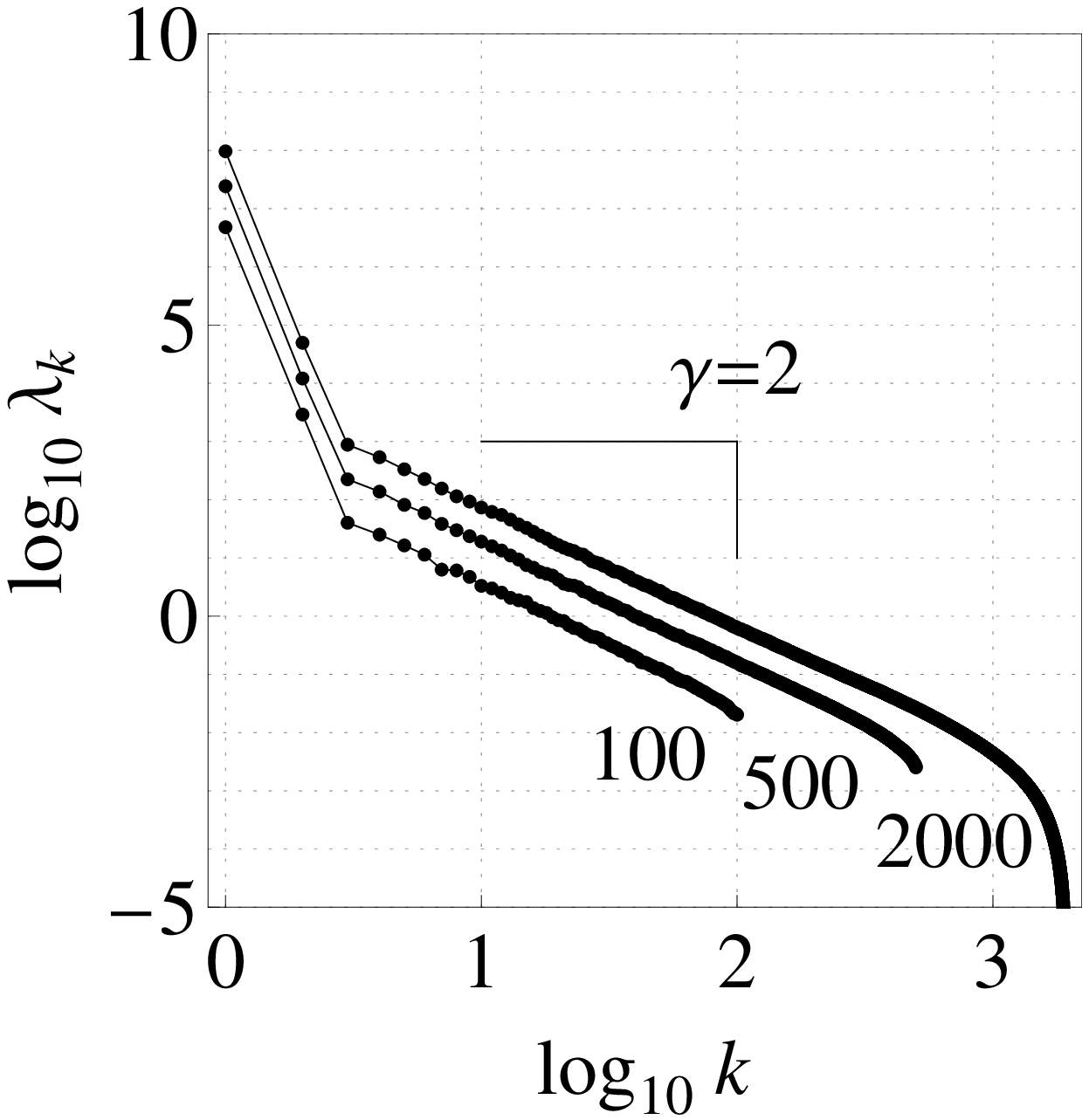}\\
%        (2a) Scree diagram (Poisson)
        \end{center}
        \end{minipage}
        %%%%%%%%%%%%%%%%%%%%%%%%%%%%%%%%%%%%%
        %%%%%%%%%%% FIG1B %%%%%%%%%%%%%%%%%%%
        \begin{minipage}{.49\linewidth}
        \begin{center}
        \includegraphics[width=.99\linewidth]{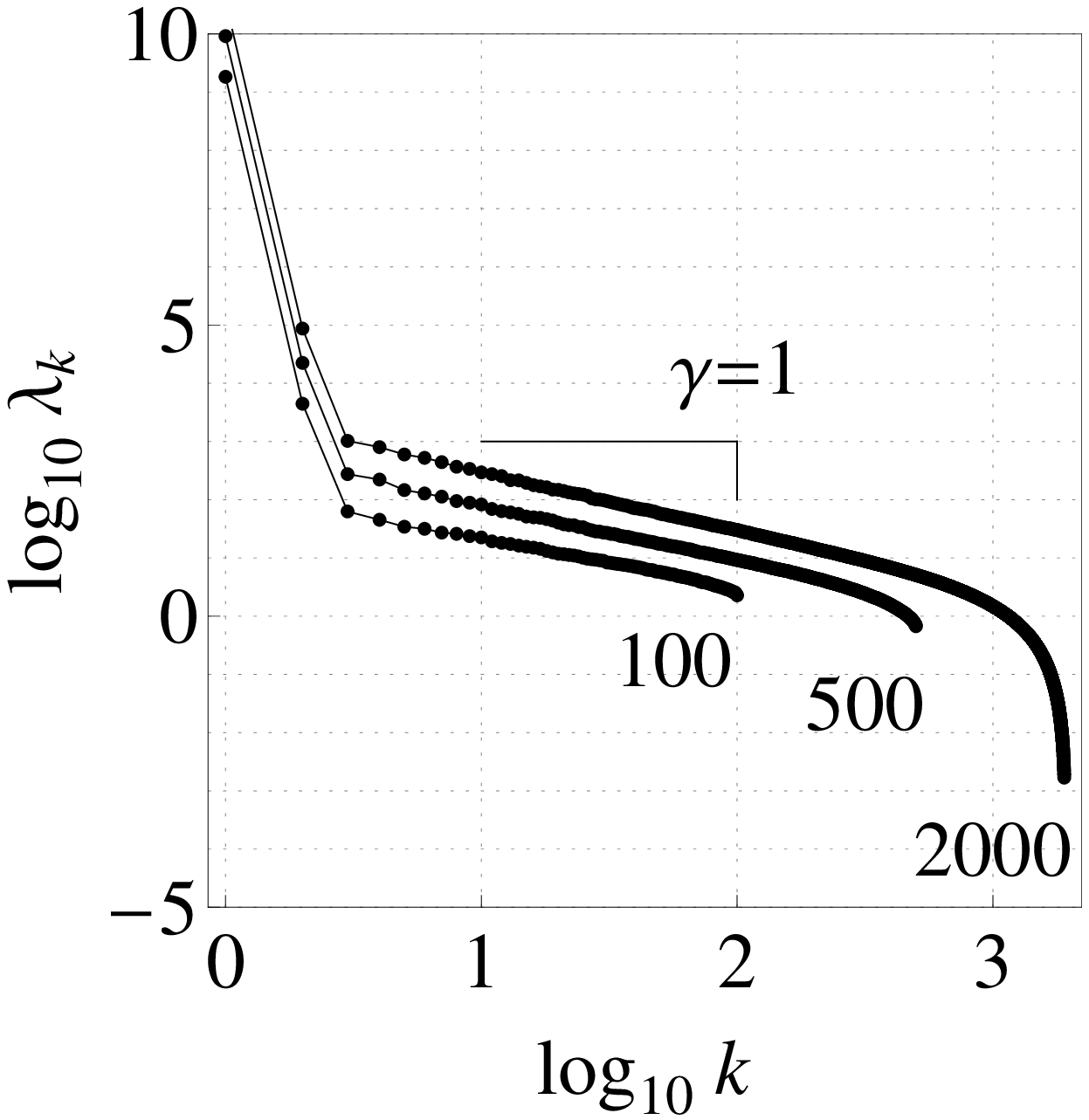}\\
%        (2b) Scree diagram (GOE)
        \end{center}
        \end{minipage}
        %%%%%%%%%%%%%%%%%%%%%%%%%%%%%%%%%%%%%
     %%%%%%%%%%%%%%%%%%%%%%%%%%%%%
     \end{minipage}     % END FRAME MINIPAGE
     %%%%%%%%%%%%%%%%%%%%%%%%%%%
     %%%%%%%%%%%%%%%%%%%%%%%%%%%%%%%%%%%%%
     \begin{minipage}{.99\linewidth}    % START FRAME MINIPAGE
     %%%%%%%%%%%%%%%%%%%%%%%%%%%%%%%%%%%%%
        %%%%%%%%%%% FIG1A %%%%%%%%%%%%%%%%%%%
        \begin{minipage}{.49\linewidth}
        \begin{center}
        \includegraphics[width=.99\linewidth]{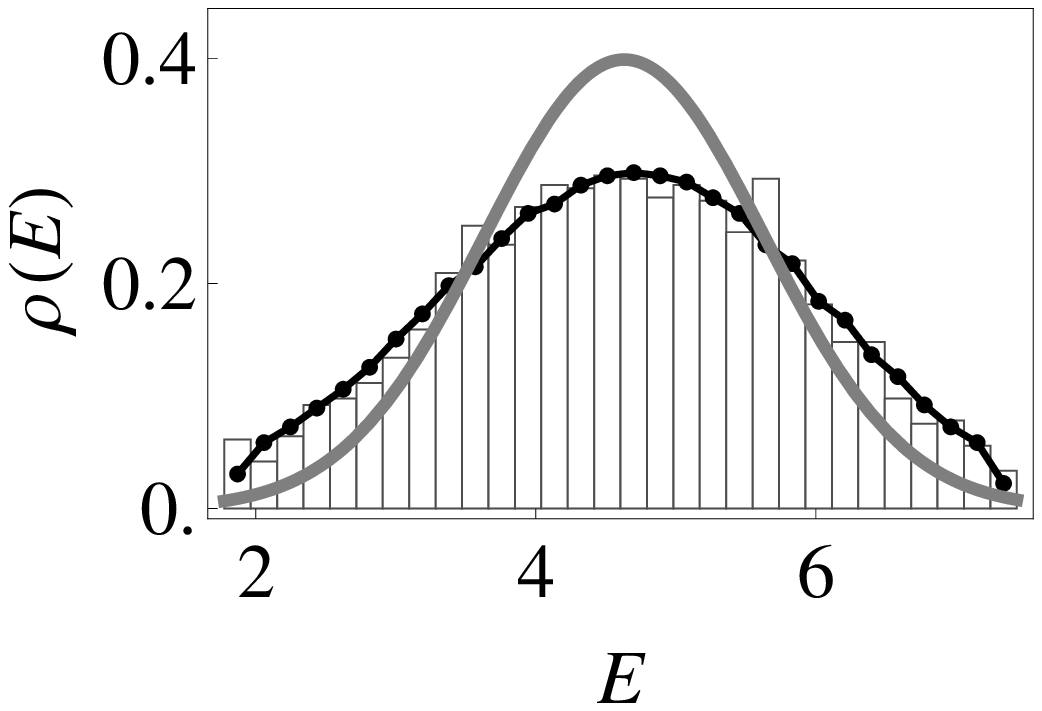}\\
%        (a) Poisson
        \end{center}
        \end{minipage}
        %%%%%%%%%%%%%%%%%%%%%%%%%%%%%%%%%%%%%
        %%%%%%%%%%% FIG1B %%%%%%%%%%%%%%%%%%%
        \begin{minipage}{.49\linewidth}
        \begin{center}
        \includegraphics[width=.99\linewidth]{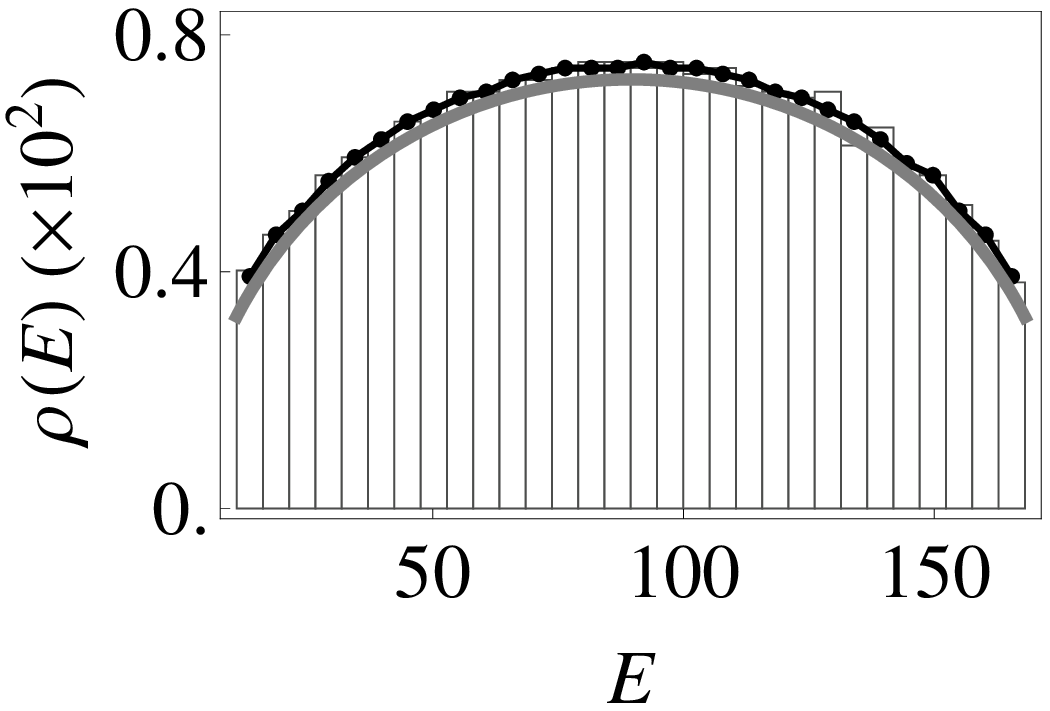}\\
%        (b) GOE
        \end{center}
        \end{minipage}
        %%%%%%%%%%%%%%%%%%%%%%%%%%%%%%%%%%%%%
     %%%%%%%%%%%%%%%%%%%%%%%%%%%%%
     \end{minipage}     % END FRAME MINIPAGE
%    %%%%%%%%%%%%%%%%%%%%%%%%%%%  
     %%%%%%%%%%%%%%%%%%%%%%%%%%%%%%%%%%%%%
     \begin{minipage}{.99\linewidth}    % START FRAME MINIPAGE
     %%%%%%%%%%%%%%%%%%%%%%%%%%%%%%%%%%%%%
        %%%%%%%%%%% FIG1A %%%%%%%%%%%%%%%%%%%
        \begin{minipage}{.49\linewidth}
        \begin{center}
        \includegraphics[width=.99\linewidth]{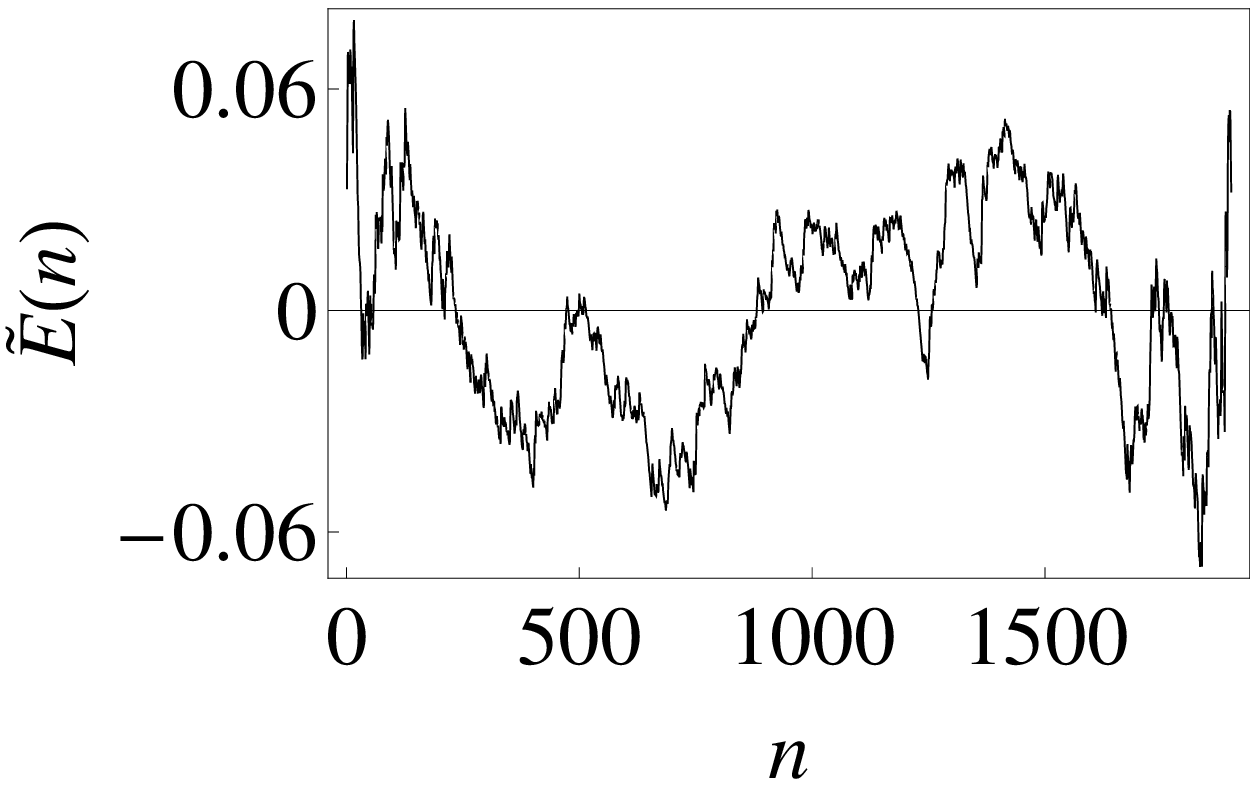}\\
%        (a) Poisson
        \end{center}
        \end{minipage}
        %%%%%%%%%%%%%%%%%%%%%%%%%%%%%%%%%%%%%
        %%%%%%%%%%% FIG1B %%%%%%%%%%%%%%%%%%%
        \begin{minipage}{.49\linewidth}
        \begin{center}
        \includegraphics[width=.99\linewidth]{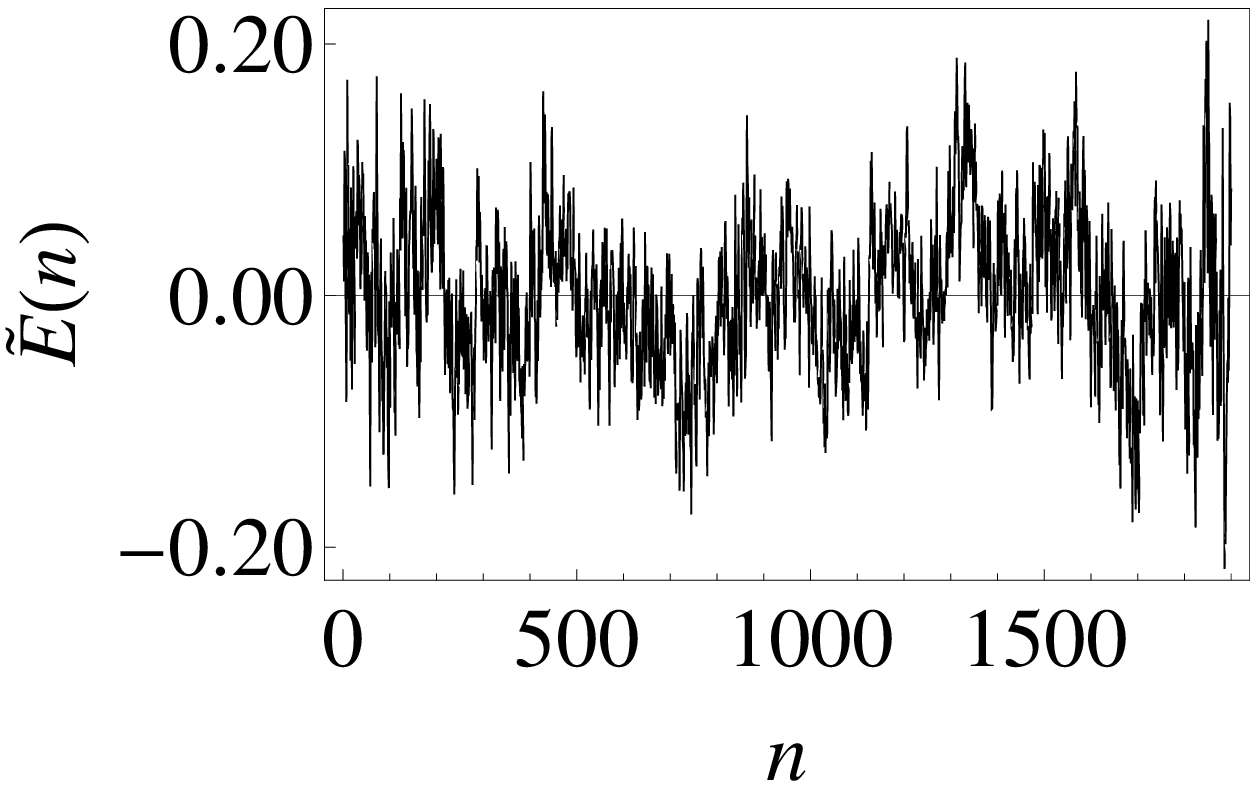}\\
%        (b) GOE
        \end{center}
        \end{minipage}
        %%%%%%%%%%%%%%%%%%%%%%%%%%%%%%%%%%%%%
     %%%%%%%%%%%%%%%%%%%%%%%%%%%%%
     \end{minipage}     % END FRAME MINIPAGE
%    %%%%%%%%%%%%%%%%%%%%%%%%%%%     
     %%%%%%%%%%%%%%%%%%%%%%%%%%%%%%%%%%%%%
     \begin{minipage}{.99\linewidth}    % START FRAME MINIPAGE
     %%%%%%%%%%%%%%%%%%%%%%%%%%%%%%%%%%%%%
        %%%%%%%%%%% FIG1A %%%%%%%%%%%%%%%%%%%
        \begin{minipage}{.49\linewidth}
        \begin{center}
        \includegraphics[width=.99\linewidth]{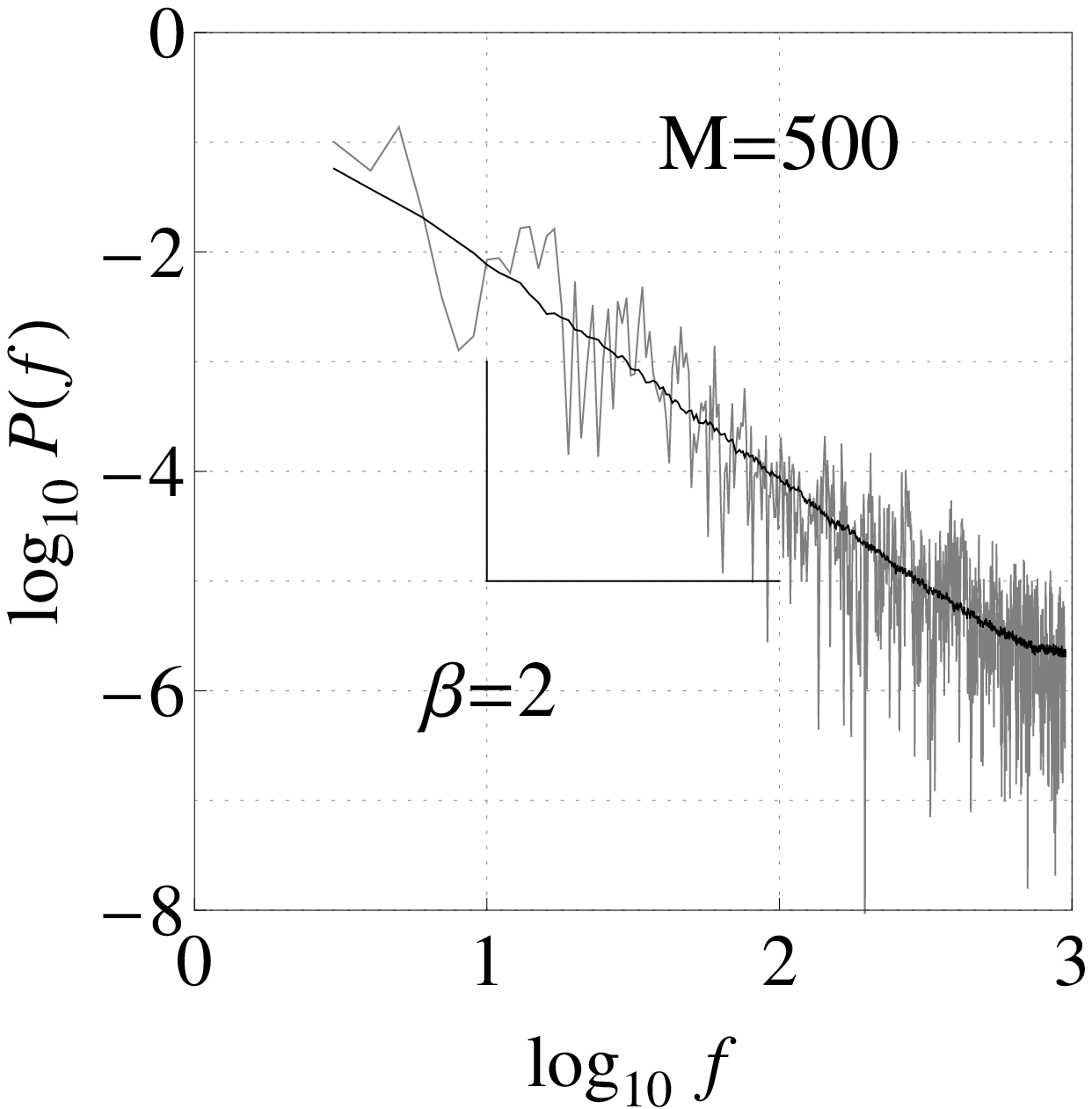}\\
        (a) Poisson
        \end{center}
        \end{minipage}
        %%%%%%%%%%%%%%%%%%%%%%%%%%%%%%%%%%%%%
        %%%%%%%%%%% FIG1B %%%%%%%%%%%%%%%%%%%
        \begin{minipage}{.49\linewidth}
        \begin{center}
        \includegraphics[width=.99\linewidth]{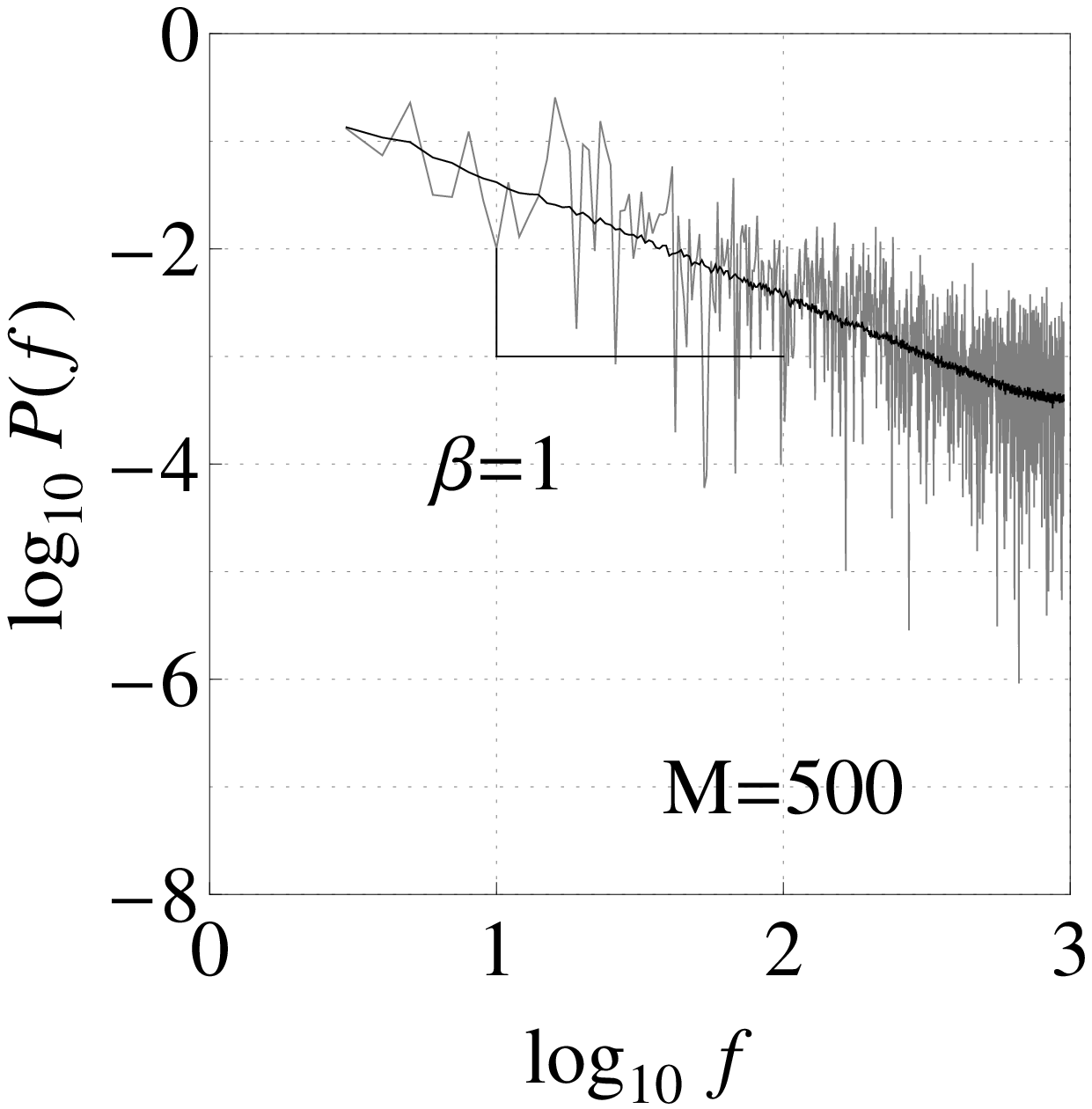}\\
        (b) GOE
        \end{center}
        \end{minipage}
        %%%%%%%%%%%%%%%%%%%%%%%%%%%%%%%%%%%%%
     %%%%%%%%%%%%%%%%%%%%%%%%%%%%%
     \end{minipage}     % END FRAME MINIPAGE
%     %%%%%%%%%%%%%%%%%%%%%%%%%%%
%%%%%%%%%%%%%%%%%%%%%%%%%%%%%%%%%%%%%
\end{minipage}
%%%%%%%%%%%%%%%%%%%%%%%%%%%%%%%%%%%%%
\caption{Results of the data-adaptive unfolding with SVD of an ensemble of $m=1 \ldots M$ spectra $E^{(m)}(n)$ with $n=1 \ldots N$ levels for the Poisson (left-hand panels (a)) and the GOE case (right-hand panels (b)), using $M=100, 500, 2000$ and $N=2000$.  (Upper row) Scree diagram of ordered partial variances $\lambda_k$, of which $\lambda_1$ and $\lambda_2$ correspond to the trend $\overline{E}^{(m)}(n)$, whereas $ \lambda_k$ with $k=3 \ldots r $ correspond to the fluctuations $\widetilde{E}^{(m)}(n)$ and follow a power law $\lambda_k \propto 1/k^\gamma$ with $\gamma=2$ (Poisson) and $\gamma=1$ (GOE). The total variance $\lambda_\mathrm{tot}=\sum_k \lambda_k$ is proportional to the ensemble size $M$. (Second row) Level density $\rho(E)$ (histogram), compared to the analytical mean density $\overline{\rho}(E)$ from the gaussian and semi-circle laws (gray line), and to the data-adaptive mean density $\rho(\overline{E})$ (black line). (Third row) Level fluctuations $\widetilde{E}^{(m)}(n)$. (Bottom row) The orresponding Fourier power spectrum follows a power law $P(f) \propto 1/f^\beta$ with $\beta=\gamma=2$ (Poisson) and $\beta=\gamma=1$ (GOE), shown for one particular spectrum realization (grey curve) and for the ensemble mean (black curve).  }
\label{FigSVD}
\end{figure}

In the present contribution, we will consider Poisson and GOE spectra ensembles. The spectra can be unfolded in a data-adaptive way applying Singular Value Decomposition (SVD) to the ensemble, and the fluctuations will be characterized by the power laws of eq.~(\ref{EqFourier}) and (\ref{EqScree}). The unfolding presented here is valid in the case of ergodic spectra, where the ensemble mean is representative for the individual spectra. More complex situations, e.g. with non-ergodic spectra, can be unfolded individually, using a variant of the present method based on Singular Spectrum Analysis (SSA). This, and other topics, such as the reconstruction of traditional fluctuation measures (NNS, $\Sigma_2$ and $\Delta_3$), and the study of transitional spectra between the Poisson and GOE limits, will be discussed elsewhere \cite{fos13}. \\

Consider an ensemble of $m=1 \ldots M$ eigenspectra $E^{(m)}(n)$, where each spectrum consists of $n=1 \ldots N$ levels. Each spectrum is conveniently accomodated in a row of the $M \times N$ dimensional matrix $\mathbf{X}$, which can now be interpreted as a multivariate time series,
\begin{equation}
\mathbf{X}=
\left(%
\begin{array}{cccc}
  E^{(1)}(1) & E^{(1)}(2) & \cdots & E^{(1)}(N) \\
  E^{(2)}(1) & E^{(2)}(2) & \cdots & E^{(2)}(N) \\
  \vdots & \vdots & \ddots & \vdots \\
  E^{(M)}(1) & E^{(M)}(2) & \cdots & E^{(M)}(N) \\
\end{array}%
\right).
\label{EqX}
\end{equation}
SVD decomposes $\mathbf{X}$ in a unique and exact way as,
\begin{equation}
\mathbf{X} = \mathbf{U} \mathbf{\Sigma} \mathbf{V}^T =  \sum_{k=1}^r \sigma_k \vec{u}_k \vec{v}_k^T, \label{EqSVD}
\end{equation}
where $\mathbf{\Sigma}$ is an $M \times N$-dimensional matrix with only diagonal elements that are the ordered singular values $\sigma_1 \geq \sigma_2 \geq \ldots \geq \sigma_r$, where $r \leq \mathrm{Min}[M,N]=\mathrm{rank}(\mathbf{X})$. The vectors $\vec{u}_k$ are orthonormal, and constitute the $k$th columns of the $M \times M$-dimensional matrix $\mathbf{U}$.  The vectors $\vec{v}_k$ are orthonormal as well, and constitute the $k$th columns of the $N \times N$-dimensional matrix $\mathbf{V}$.  $\vec{u}_k \vec{v}_k^T \equiv \vec{u}_k \otimes \vec{v}_k$ is the outer product of  $\vec{u}_k$ and $\vec{v}_k$. In this way, any matrix row of $\mathbf{X}$ containing a particular excitation spectrum $E^{(m)}(n)$ can be written as, 
\begin{equation}
E^{(m)}(n) =  \sum_{k=1}^r \sigma_k U_{mk} \vec{v}^T_k(n),
\label{EqDecomposition}
\end{equation}
and can be interpreted as a superposition of basis vectors $\vec{v}_k$, that are common for the whole ensemble $\mathbf{X}$, and where the matrix elements $U_{mk}$ serve as coefficients. On the other hand, the singular values $\sigma_k$ can be interpreted as weights that distinguish between trend and fluctuation components. A spectrum is a monotonous function that has a dominant trend, with superposed fluctuations that are typically orders of magnitude smaller. Consequently, the variability of a spectrum will be due principally to its trend components, characterized by very large partial variances $\lambda_k=\sigma_k^2$, whereas the fluctuation components will be associated with much smaller partial variances. Thus, in eq.~(\ref{EqDecomposition}), we expect to be able to separate in a data-adaptive way the trend $\overline{E}^{(m)}(n)$ from the fluctuations $\widetilde{E}^{(m)}(n)$ in the following way, 
\begin{equation}
\left\{
\begin{array}{c}
  \overline{E}^{(m)}(n) =  \sum_{k=1}^{n_T} \sigma_k U_{mk} \vec{v}_k^T(n) \\
  \widetilde{E}^{(m)}(n) =  \sum_{k=n_T+1}^r \sigma_k U_{mk} \vec{v}_k^T(n), \\
\end{array}
\right.
\label{EqTrendFluct}
\end{equation}
where $n_T$ is the number of components to be included in the trend (excluded from the fluctuations).

\begin{figure}[htb!]
%%%%%%%%%%%%%%%%%%%%%%%%%%%%%%%%%%%%%
\begin{minipage}{.99\linewidth}    % START FRAME MINIPAGE
%%%%%%%%%%%%%%%%%%%%%%%%%%%%%%%%%%%%%
     %%%%%%%%%%%%%%%%%%%%%%%%%%%%%%%%%%%%%
     \begin{minipage}{.99\linewidth}    % START FRAME MINIPAGE
     %%%%%%%%%%%%%%%%%%%%%%%%%%%%%%%%%%%%%
        %%%%%%%%%%% FIG1A %%%%%%%%%%%%%%%%%%%
        \begin{minipage}[t]{.24\linewidth}
        \begin{center}
        \includegraphics[width=.99\linewidth]{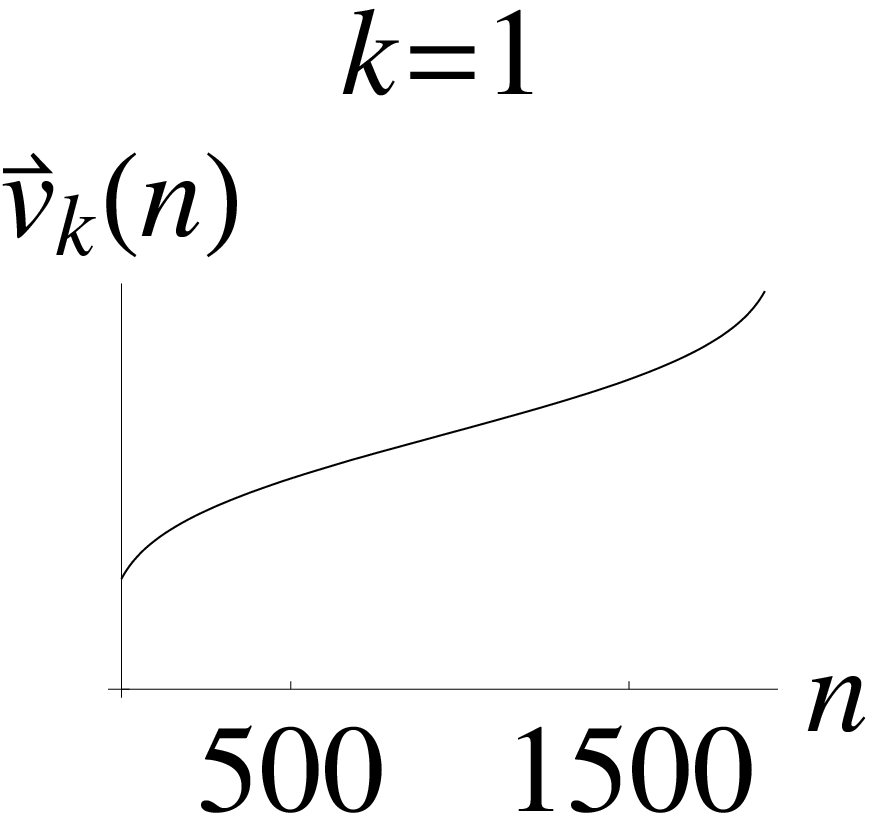}\\
%        (2a) Scree diagram (Poisson)
        \end{center}
        \end{minipage}
        %%%%%%%%%%%%%%%%%%%%%%%%%%%%%%%%%%%%%
        %%%%%%%%%%% FIG1B %%%%%%%%%%%%%%%%%%%
        \begin{minipage}[t]{.24\linewidth}
        \begin{center}
        \includegraphics[width=.99\linewidth]{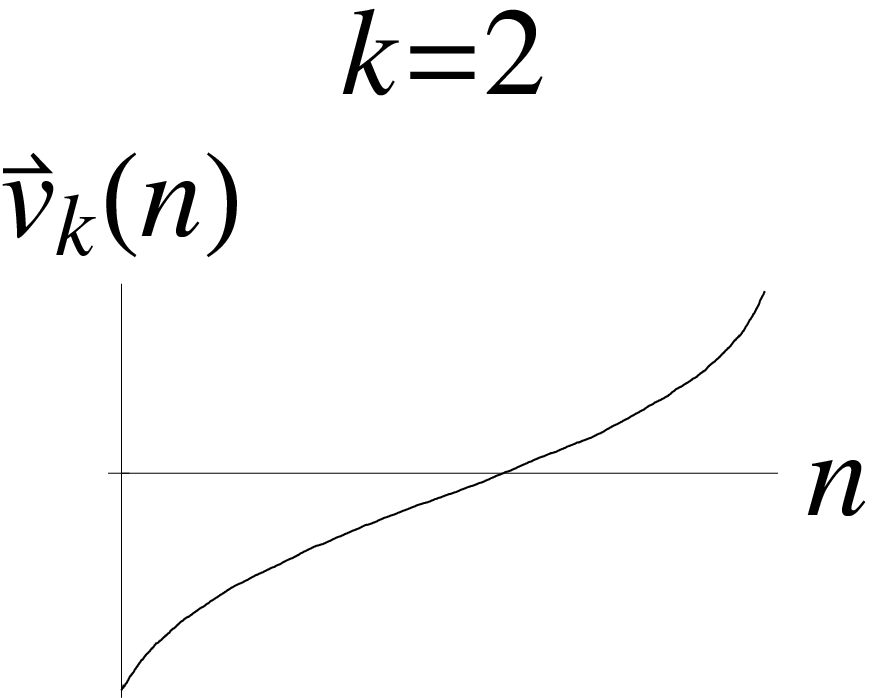}\\
%        (2b) Scree diagram (GOE)
        \end{center}
        \end{minipage}
        %%%%%%%%%%%%%%%%%%%%%%%%%%%%%%%%%%%%%
        %%%%%%%%%%% FIG1B %%%%%%%%%%%%%%%%%%%
        \begin{minipage}[t]{.24\linewidth}
        \begin{center}
        \includegraphics[width=.99\linewidth]{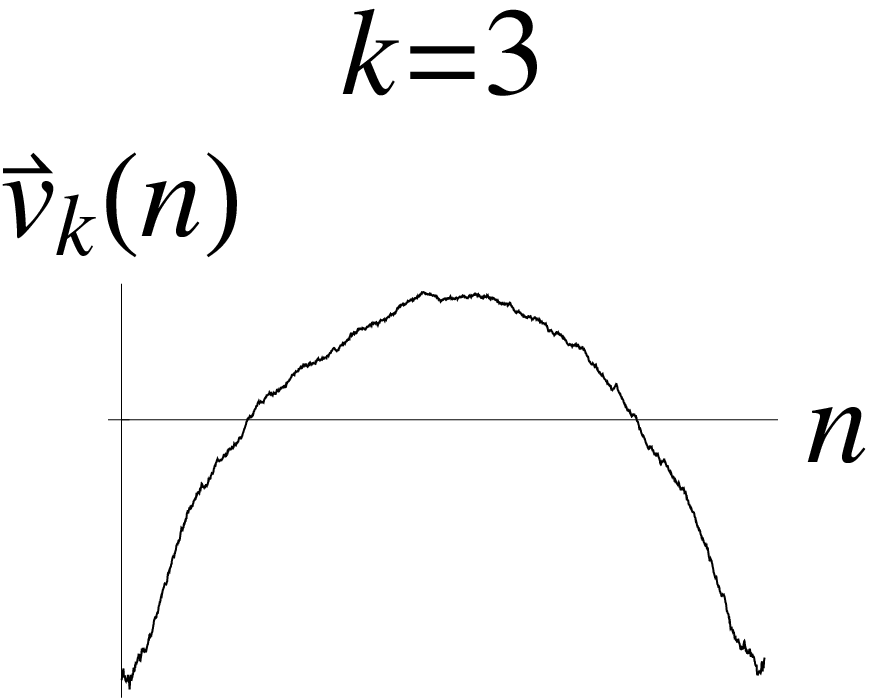}\\
%        (2b) Scree diagram (GOE)
        \end{center}
        \end{minipage}
        %%%%%%%%%%%%%%%%%%%%%%%%%%%%%%%%%%%%%
        %%%%%%%%%%% FIG1B %%%%%%%%%%%%%%%%%%%
        \begin{minipage}[t]{.24\linewidth}
        \begin{center}
        \includegraphics[width=.99\linewidth]{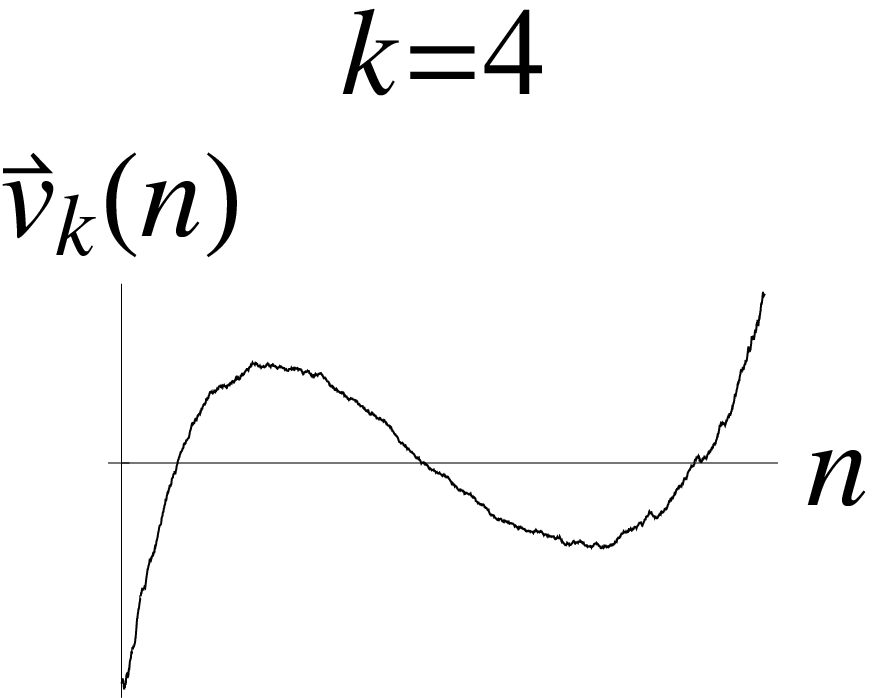}\\
%        (2b) Scree diagram (GOE)
        \end{center}
        \end{minipage}
        %%%%%%%%%%%%%%%%%%%%%%%%%%%%%%%%%%%%%
     %%%%%%%%%%%%%%%%%%%%%%%%%%%%%
     \end{minipage}     % END FRAME MINIPAGE
     %%%%%%%%%%%%%%%%%%%%%%%%%%%
     %%%%%%%%%%%%%%%%%%%%%%%%%%%%%%%%%%%%%
     \begin{minipage}{.99\linewidth}    % START FRAME MINIPAGE
     %%%%%%%%%%%%%%%%%%%%%%%%%%%%%%%%%%%%%
        %%%%%%%%%%% FIG1A %%%%%%%%%%%%%%%%%%%
        \begin{minipage}[t]{.24\linewidth}
        \begin{center}
        \includegraphics[width=.99\linewidth]{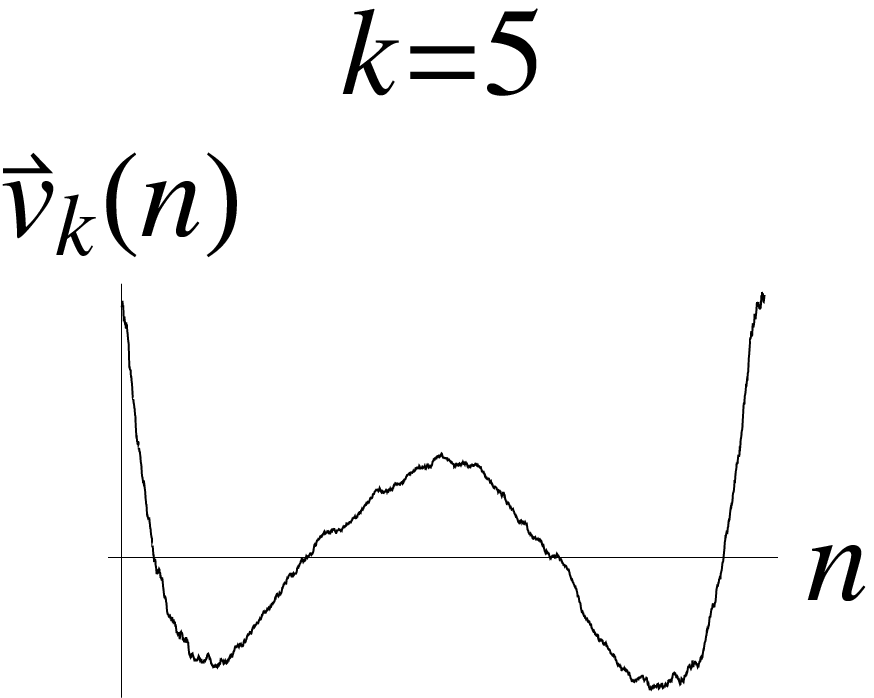}\\
%        (2a) Scree diagram (Poisson)
        \end{center}
        \end{minipage}
        %%%%%%%%%%%%%%%%%%%%%%%%%%%%%%%%%%%%%
        %%%%%%%%%%% FIG1B %%%%%%%%%%%%%%%%%%%
        \begin{minipage}[t]{.24\linewidth}
        \begin{center}
        \includegraphics[width=.99\linewidth]{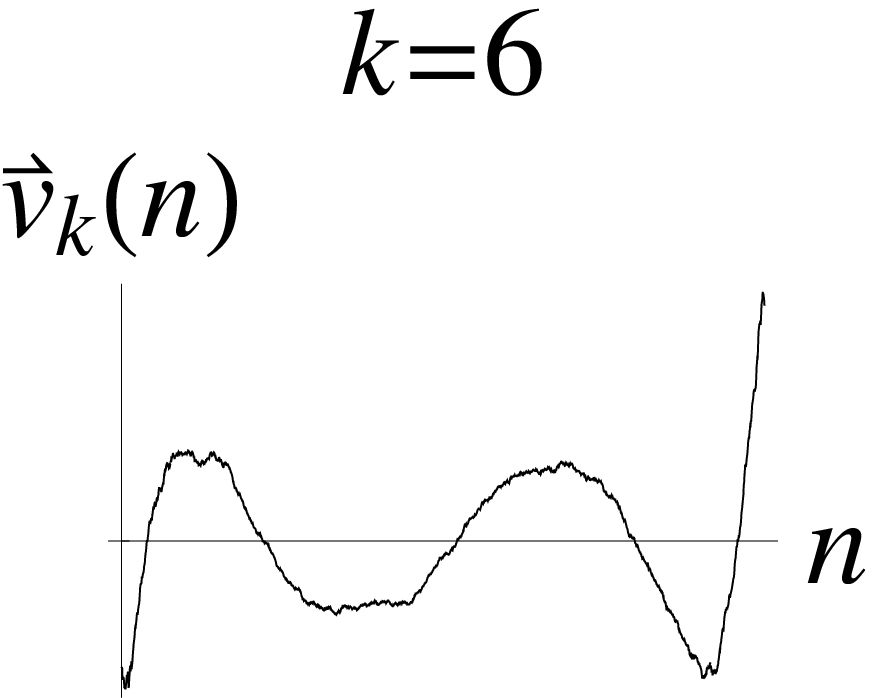}\\
%        (2b) Scree diagram (GOE)
        \end{center}
        \end{minipage}
        %%%%%%%%%%%%%%%%%%%%%%%%%%%%%%%%%%%%%
        %%%%%%%%%%% FIG1B %%%%%%%%%%%%%%%%%%%
        \begin{minipage}[t]{.24\linewidth}
        \begin{center}
        \includegraphics[width=.99\linewidth]{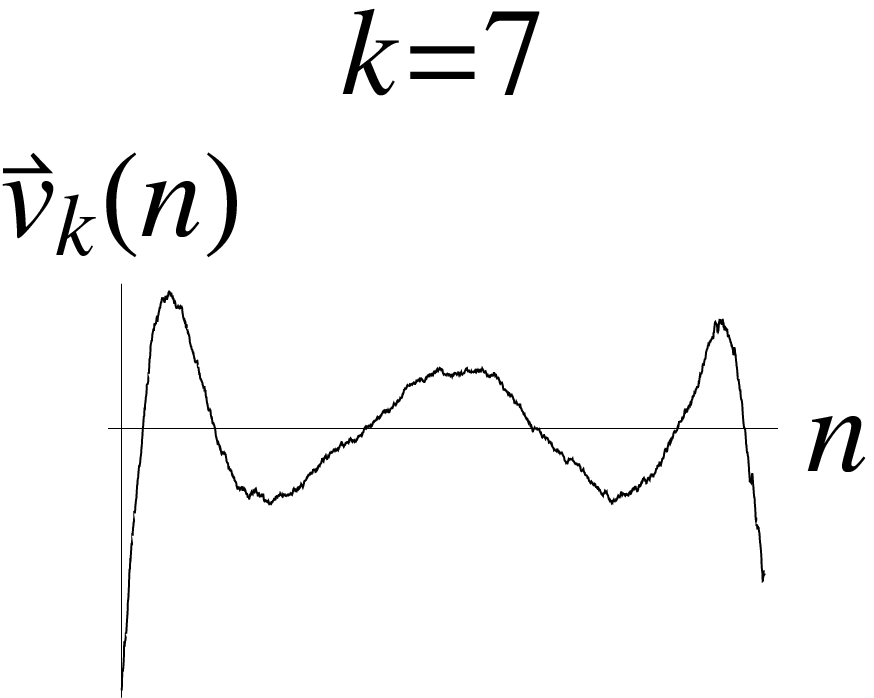}\\
%        (2b) Scree diagram (GOE)
        \end{center}
        \end{minipage}
        %%%%%%%%%%%%%%%%%%%%%%%%%%%%%%%%%%%%%
        %%%%%%%%%%% FIG1B %%%%%%%%%%%%%%%%%%%
        \begin{minipage}[t]{.24\linewidth}
        \begin{center}
        \includegraphics[width=.99\linewidth]{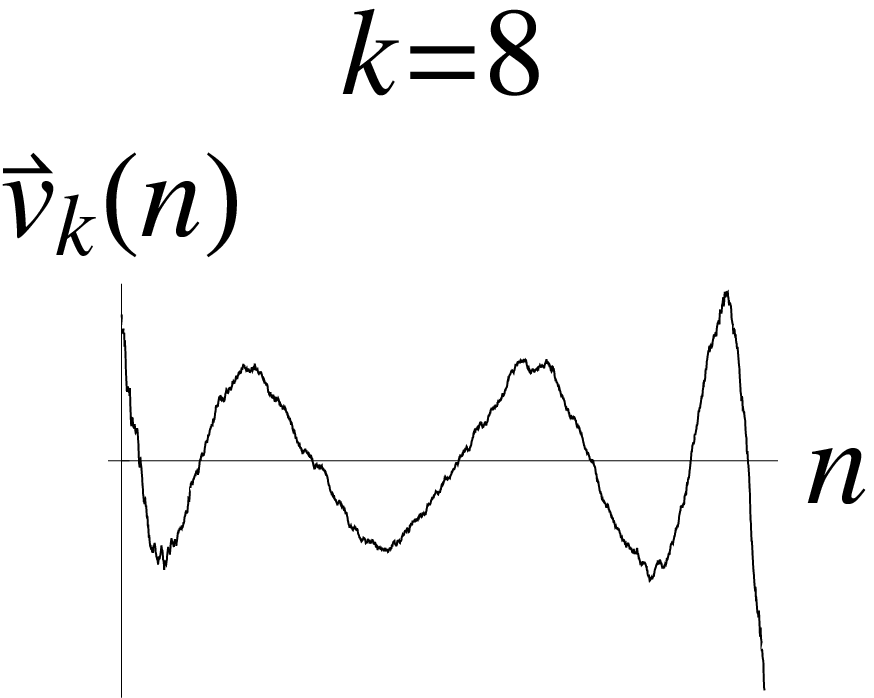}\\
%        (2b) Scree diagram (GOE)
        \end{center}
        \end{minipage}
        %%%%%%%%%%%%%%%%%%%%%%%%%%%%%%%%%%%%%
     %%%%%%%%%%%%%%%%%%%%%%%%%%%%%
     \end{minipage}     % END FRAME MINIPAGE
     %%%%%%%%%%%%%%%%%%%%%%%%%%%
     (a) Poisson \vspace{0.5cm}\\ 
     %%%%%%%%%%%%%%%%%%%%%%%%%%%%%%%%%%%%%
     \begin{minipage}{.99\linewidth}    % START FRAME MINIPAGE
     %%%%%%%%%%%%%%%%%%%%%%%%%%%%%%%%%%%%%
        %%%%%%%%%%% FIG1A %%%%%%%%%%%%%%%%%%%
        \begin{minipage}[t]{.24\linewidth}
        \begin{center}
        \includegraphics[width=.99\linewidth]{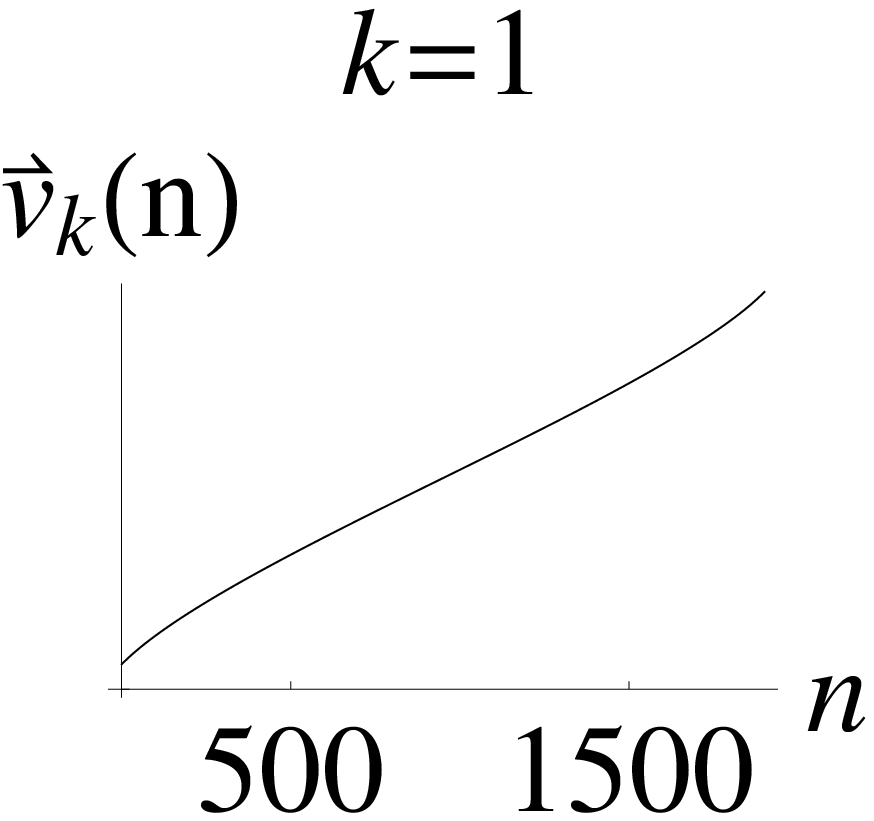}\\
%        (2a) Scree diagram (Poisson)
        \end{center}
        \end{minipage}
        %%%%%%%%%%%%%%%%%%%%%%%%%%%%%%%%%%%%%
        %%%%%%%%%%% FIG1B %%%%%%%%%%%%%%%%%%%
        \begin{minipage}[t]{.24\linewidth}
        \begin{center}
        \includegraphics[width=.99\linewidth]{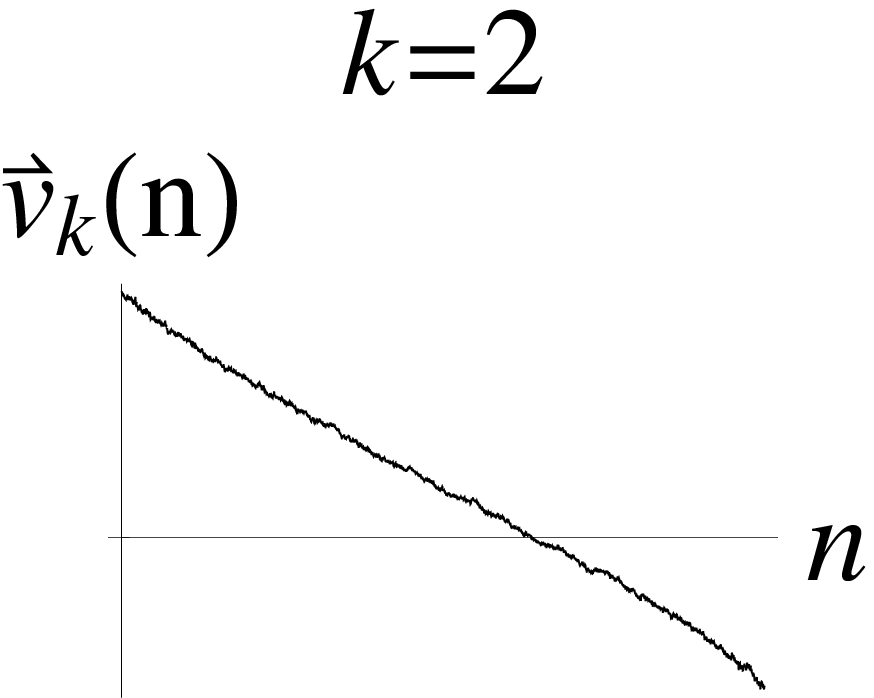}\\
%        (2b) Scree diagram (GOE)
        \end{center}
        \end{minipage}
        %%%%%%%%%%%%%%%%%%%%%%%%%%%%%%%%%%%%%
        %%%%%%%%%%% FIG1B %%%%%%%%%%%%%%%%%%%
        \begin{minipage}[t]{.24\linewidth}
        \begin{center}
        \includegraphics[width=.99\linewidth]{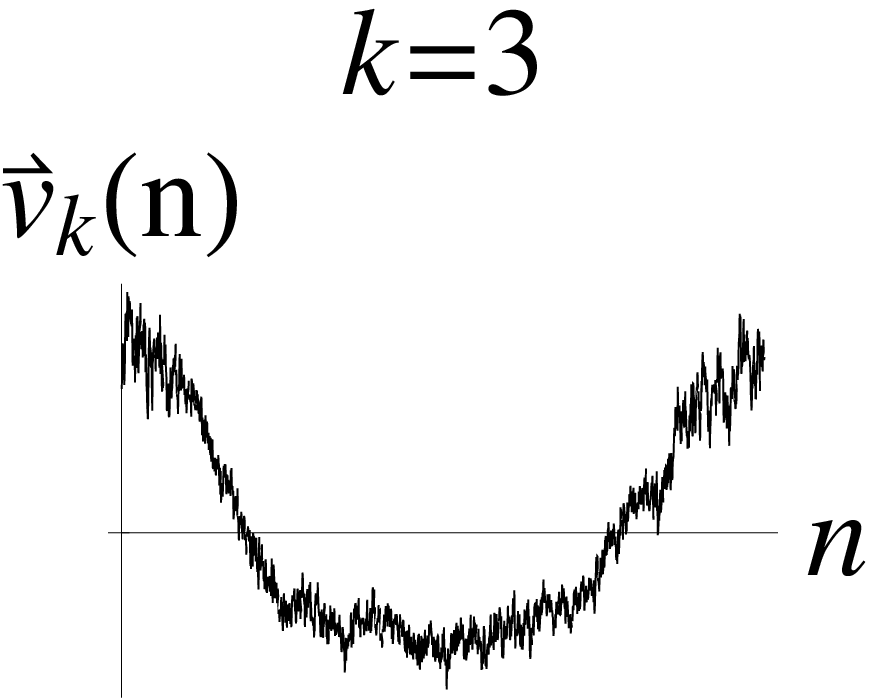}\\
%        (2b) Scree diagram (GOE)
        \end{center}
        \end{minipage}
        %%%%%%%%%%%%%%%%%%%%%%%%%%%%%%%%%%%%%
        %%%%%%%%%%% FIG1B %%%%%%%%%%%%%%%%%%%
        \begin{minipage}[t]{.24\linewidth}
        \begin{center}
        \includegraphics[width=.99\linewidth]{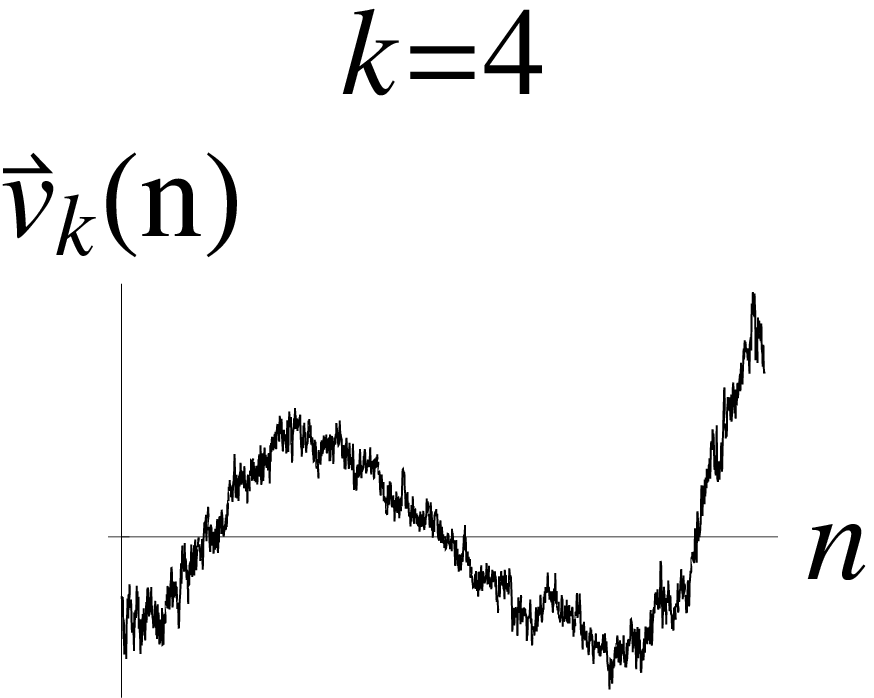}\\
%        (2b) Scree diagram (GOE)
        \end{center}
        \end{minipage}
        %%%%%%%%%%%%%%%%%%%%%%%%%%%%%%%%%%%%%
     %%%%%%%%%%%%%%%%%%%%%%%%%%%%%
     \end{minipage}     % END FRAME MINIPAGE
     %%%%%%%%%%%%%%%%%%%%%%%%%%%
     %%%%%%%%%%%%%%%%%%%%%%%%%%%%%%%%%%%%%
     \begin{minipage}{.99\linewidth}    % START FRAME MINIPAGE
     %%%%%%%%%%%%%%%%%%%%%%%%%%%%%%%%%%%%%
        %%%%%%%%%%% FIG1A %%%%%%%%%%%%%%%%%%%
        \begin{minipage}[t]{.24\linewidth}
        \begin{center}
        \includegraphics[width=.99\linewidth]{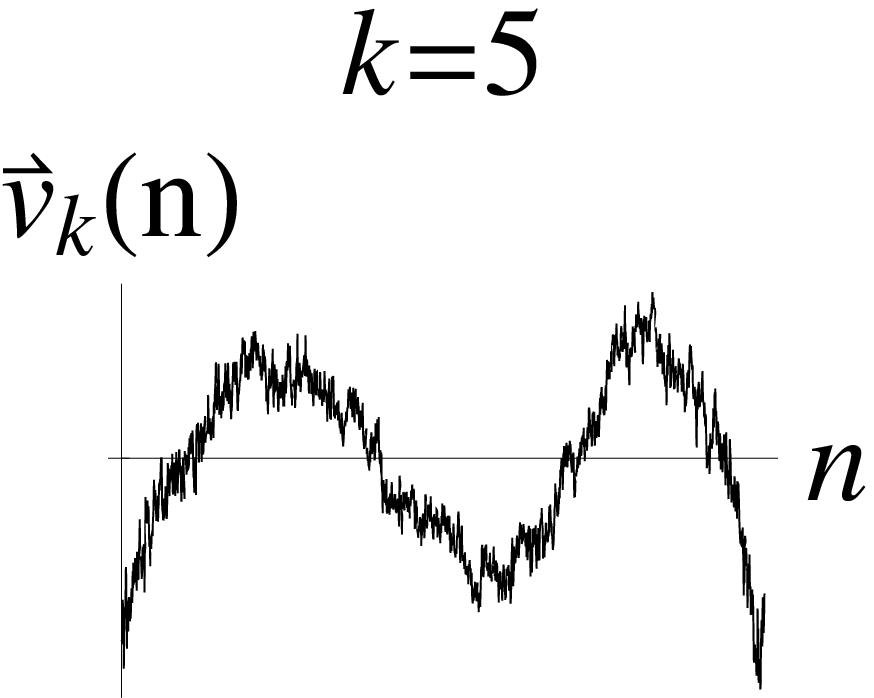}\\
%        (2a) Scree diagram (Poisson)
        \end{center}
        \end{minipage}
        %%%%%%%%%%%%%%%%%%%%%%%%%%%%%%%%%%%%%
        %%%%%%%%%%% FIG1B %%%%%%%%%%%%%%%%%%%
        \begin{minipage}[t]{.24\linewidth}
        \begin{center}
        \includegraphics[width=.99\linewidth]{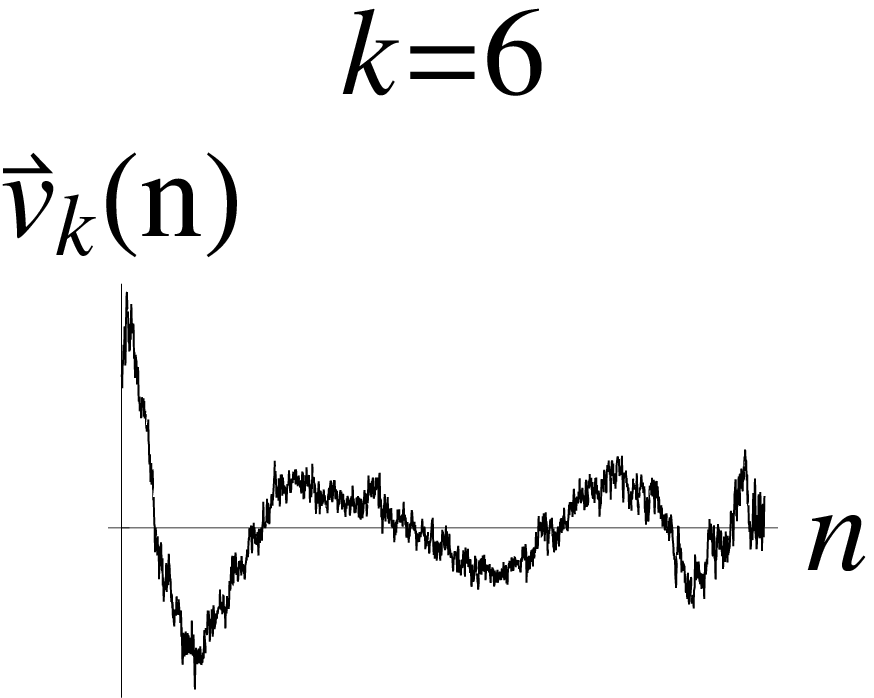}\\
%        (2b) Scree diagram (GOE)
        \end{center}
        \end{minipage}
        %%%%%%%%%%%%%%%%%%%%%%%%%%%%%%%%%%%%%
        %%%%%%%%%%% FIG1B %%%%%%%%%%%%%%%%%%%
        \begin{minipage}[t]{.24\linewidth}
        \begin{center}
        \includegraphics[width=.99\linewidth]{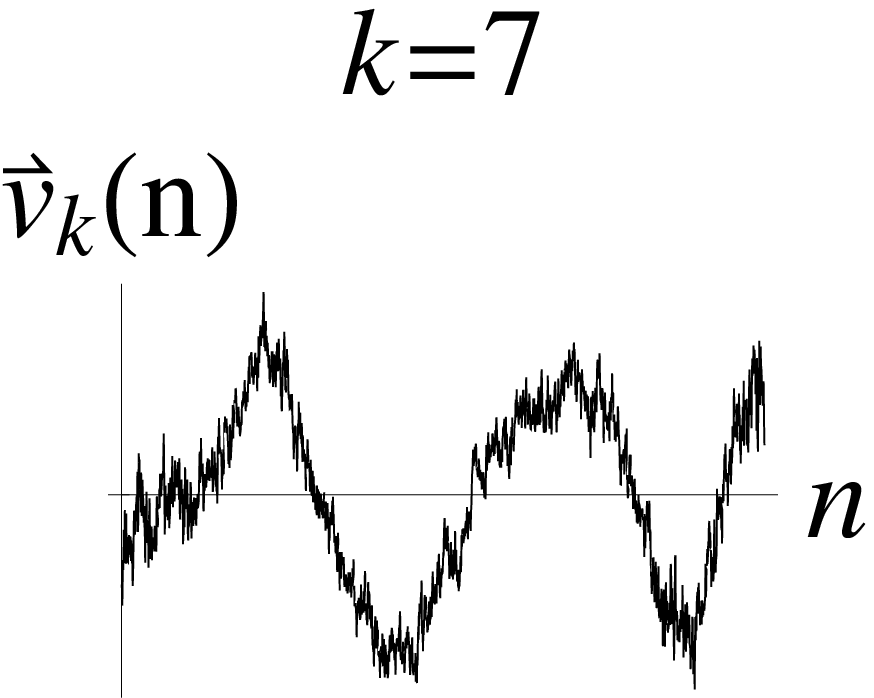}\\
%        (2b) Scree diagram (GOE)
        \end{center}
        \end{minipage}
        %%%%%%%%%%%%%%%%%%%%%%%%%%%%%%%%%%%%%
        %%%%%%%%%%% FIG1B %%%%%%%%%%%%%%%%%%%
        \begin{minipage}[t]{.24\linewidth}
        \begin{center}
        \includegraphics[width=.99\linewidth]{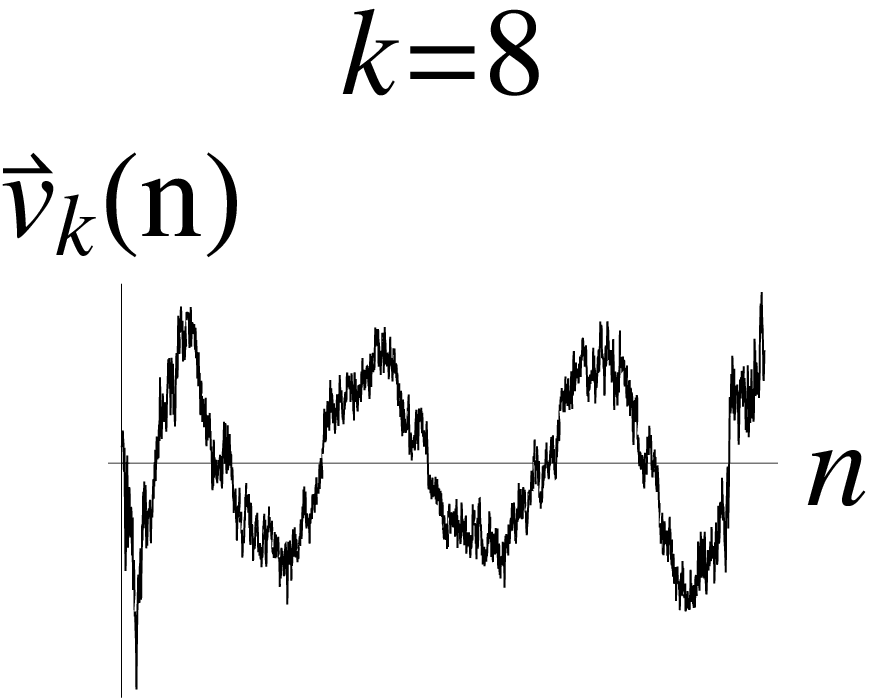}\\
%        (2b) Scree diagram (GOE)
        \end{center}
        \end{minipage}
        %%%%%%%%%%%%%%%%%%%%%%%%%%%%%%%%%%%%%
     %%%%%%%%%%%%%%%%%%%%%%%%%%%%%
     \end{minipage}     % END FRAME MINIPAGE
     %%%%%%%%%%%%%%%%%%%%%%%%%%%     
     (b) GOE \\ 
%%%%%%%%%%%%%%%%%%%%%%%%%%%%%%%%%%%%%
\end{minipage}
%%%%%%%%%%%%%%%%%%%%%%%%%%%%%%%%%%%%%
\caption{The vectors $\vec{v}_k$ constitute an orthonormal basis for the ensemble $\mathbf{X}$ of eq~(\ref{EqX}). The first 8 vectors $\vec{v}_k$ are shown for ensemble size $M=500$, (a) for the Poisson case, and (b) for the GOE case. Vectors $\vec{v}_1$ and $\vec{v}_2$ are monotonous and serve as a basis for the trend $\overline{E}^{(m)}(n)$ of all realizations $m=1 \ldots M$ of the ensemble. The higher-order vectors $\vec{v}_k$ with $k=3 \ldots r$ oscillate and serve as a basis for the fluctuations $\widetilde{E}^{(m)}(n)$. }
\label{FigNormalModes}
\end{figure}

In Fig.~\ref{FigSVD}, results are shown for SVD applied to Poisson and GOE ensembles with $M=100, 500, 2000$ realizations $E^{(m)}(n)$, where each spectrum contains $N=2000$ levels. To take into account only the central part of the spectrum (within 2 standard deviations), $2.5\%$ of the lower and upper levels were discarded. From the scree diagram of ordered partial variances follows that $\lambda_1$ and $\lambda_2$ are orders of magnitude larger than the other partial variances, and that they are responsible for the major part of the total variance $\lambda_\mathrm{tot}=\sum_k \lambda_k$, both in the Poisson as in the GOE case. In Fig.~\ref{FigNormalModes}, it can be seen that the associated basis vectors $\vec{v}_1$ and $\vec{v}_2$ behave monotonically. Based on these arguments, it can be concluded that the first two vectors constitute the basis states for the trend $\overline{E}^{(m)}(n)$ of each of the realizations of the Poisson and the GOE ensemble, see eq.~(\ref{EqTrendFluct}). On the other hand, the larger-order partial variances $\lambda_k$ with $3 \leq k \leq r$ behave as the power law of eq.~(\ref{EqScree}) with $\gamma \approx 2$ (Poisson) and $\gamma \approx 1$ (GOE), so that already during the unfolding procedure one distinguishes between the two cases. The vectors $\vec{v}_k$ with $3 \leq k \leq r$, associated to the larger-order partial variances, oscillate, and they constitute the basis vectors for the fluctuations $\widetilde{E}^{(m)}(n)$, see eq.~(\ref{EqTrendFluct}). It can be appreciated that the more realizations $M$ the ensemble contains, the larger the total variance $\lambda_\mathrm{tot}$ of the ensemble becomes ($\lambda_\mathrm{tot} \propto M$), and the larger the number $r$ of components with which each spectrum is decomposed. Because of the power-law $\lambda_k \propto 1/k^\gamma$ of eq.~(\ref{EqScree}), higher-order $\lambda_k$ will contribute less to the total variance, and in order for the behaviour of the scree diagram to be independent from the ensemble size, the individual partial variances $\lambda_k$ must also grow with $M$. In the following, results will be presented for intermediate ensemble sizes of $M=500$, because for small $M$ the range of the power law of eq.~(\ref{EqScree}) is reduced, whereas for very large $M$ the basis $\left\{ \vec{v}_k, k=1 \ldots r \right\}$ can become overcomplete, leading to a tail of insignificantly small partial variances $\lambda_k$ in the scree diagram. However, the statistical results are independent of the particular choice of $M$. Note that the vectors $\vec{v}_k$ with $k \geq 3$ correspond with the fluctuation \emph{normal modes} of ref.~\cite{and99,jac01}, that were obtained after a prior and separate unfolding step. In the present contribution, both the trend basis vectors $\vec{v}_1, \vec{v}_2$ and the fluctuation basis vectors $\vec{v}_k$ ($k \geq 3$) are obtained during the data-adaptive unfolding itself. There is no formal difference between trend and fluctuation basis vectors, other than the former vectors behaving monotonically. Also in \cite{jac01}, it was stated that an appropriate unfolding procedure should reflect the spectral scale which is relevant for the physical properties in question, and that such a scale is not always apparent given the usual ad hoc treatment of unfolding. In the context of the determination of the trend of non-stationary time series, it is known that without a reference to a particular scale the trend will be confusingly mixed with the local fluctuations \cite{wu07}. In the present contribution, the different scales of the trend and fluctuations modes follow directly from the scree diagram of ordered partial variances $\lambda_k$. 

\vspace{0.5cm}

Also in Fig.~\ref{FigSVD}, results are shown for one particular realization of a Poisson and a GOE spectrum, for the mean level density $\overline{\rho}(E)$, the fluctuations $\widetilde{E}^{(m)}(n)$, and the corresponding Fourier power spectrum $P(f)$, after the data-adaptive separation of trend and fluctuation modes as described above. It can be seen that in the GOE case, the global level density $\overline{\rho}(E)$ has converged to the asymptotic semi-circle law. On the other hand, in the Poisson case, the Gaussian distribution does not describe well the global level density. Thus, although an analytical formula is known to describe the global level density in the asymptotic case, here, it can not be applied to perform the unfolding of the spectrum of the present finite matrix. On the other hand, the global level density $\overline{\rho}(E)$ can be determined in a data-adaptive way as the density $\rho(\overline{E})$ of the smooth trend approximation $\overline{E}^{(m)}(n)$ to a specific spectrum of interest. It can be seen that $\rho(\overline{E})$ describes well the global level density in both the Poisson and the GOE case. Next, the level fluctuations $\widetilde{E}^{(m)}(n)$ of one particular Poisson and GOE spectrum are shown, according to eq.~(\ref{EqTrendFluct}), with $n_T=2$ trend components as clearly follows from the scree diagram. Finally, the Fourier power spectrum is presented for the level fluctuations shown. It can be seen that the power spectrum obeys the power law of eq.~(\ref{EqFourier}) with $\beta=\gamma=2$ in the Poisson case, and $\beta=\gamma=1$ in the GOE case. This power law is even more apparent if the power spectrum is averaged over all realizations $m=1 \ldots M$ of the ensemble. Note that near the maximum frequency $f=N/2$ (Nyquist frequency), there is a deviation from the power law, as previously described in ref.~\cite{fal04}. 

\vspace{0.5cm}

In conclusion, we presented a method to perform the unfolding of random-matrix spectra in a data-adaptive way. The unfolding is a long-standing problem in the field of Random Matrix Theory (RMT), and the complications of the unfolding technique have become topical again due to the recent spread of RMT techniques to areas as diverse as the study of eigenspectra from adjacency matrices in networks, and correlation matrices in finance, the climate, magneto- and electroencephalography, etc. In the present contribution, in the first place, we suggested to interpret a matrix eigenspectrum directly as a time series and to apply techniques from signal analysis to perform the unfolding procedure of separation of trend and fluctuation components in a data-adaptive way. Secondly, we proposed one particular method, based on Singular Value Decomposition (SVD) with which this unfolding can be realized. We applied the method to ensembles of Poisson and GOE spectra. Already during the unfolding procedure, a power law is obtained for the fluctuations that distinguishes between the Poisson and the GOE case. Such a data-adaptive unfolding should be general enough to be applicable as well to spectra with other symmetries. 

\vspace{0.5cm}

We acknowledge financial support from CONACYT (CB-2011-01-167441, CB-2010-01-155663, I010/266/2011/C-410-11) and PAPIIT-DGAPA (IN114411). This work was partly funded by the European project FP7-PEOPLE-2009-IRSES-247541-MATSIQEL and the Instituto Nacional de Geriatr\'{i}a (project DI-PI-002/2012). The authors wish to thank A. Frank and collaborators for fruitful discussions.

\end{document}